\newcommand{\beq}{\begin{equation}}
\newcommand{\eeq}{\end{equation}}
\newcommand{\beqa}{\begin{eqnarray}}
\newcommand{\eeqa}{\end{eqnarray}}

\input{aipcheck}

\documentclass[
    ,final            % use final for the camera ready runs
  ]
  {aipproc}

\layoutstyle{6x9}

\usepackage{amsmath}
\usepackage{amssymb}
\usepackage{amsfonts}
\usepackage{amstext}
\usepackage{amsbsy}
\usepackage[mathscr]{eucal}
\usepackage{graphicx}
\usepackage{color}
\usepackage[all]{xy}

\begin{document}

\title{Strategies in a Symmetric Quantum Kolkata Restaurant Problem}


\author{Puya Sharif and Hoshang Heydari}{
  address={Physics Department, Stockholm University 10691 Stockholm, Sweden \newline E-mail:ps@puyasharif.net }
}

\begin{abstract}
The Quantum Kolkata restaurant problem is a multiple-choice version of the quantum minority game, where a set of $n$ non-communicating players have to chose between one of $m$ choices. A payoff is granted to the players that make a unique choice. It has previously been shown that shared entanglement and quantum operations can aid the players to coordinate their actions and acquire higher payoffs than is possible with classical randomization. In this paper the initial quantum state is expanded to a family of GHZ-type states and strategies are discussed in terms of possible final outcomes. It is shown that the players individually seek outcomes that maximize the collective good.
\end{abstract}

\maketitle

\section{Introduction}

\subsection{Games and framework}
Game theory is the study of systematic and strategic decision-making in interactive situations of conflict and cooperation. The models are widely
used in economics, political science, biology and computer science to capture the behavior of
individual participants in terms of responses to strategies of the rest. The field attempts to describe how decision makers
do and should interact within a well-defined system of rules to maximize their satisfaction with the outcome \cite{GT-Critical,GT-fudenberg,Course in GT}. A game is a model of the strategies of these decision makers or players as we will call them, in terms of choices made by each of them. They are assumed to all have individual preference profiles $\sigma_{x_1}\succeq \sigma_{x_2}\succeq \cdots \succeq \sigma_{x_m}$ over a set of $m$ outcomes $\{\sigma_{x_j}\}$, where $'\succeq\,'$ should be interpreted as "preferred by", and the $x_j$'s as indices for possible outcomes. An outcome or strategy profile $\sigma \in S_n \times S_{n-1} \times \cdots \times S_1 = S$ is equivalent to the combination of the strategies $s^i_j \in S_i$ of the participants, where $s^i_j$ is the $j$'th strategy of player $i$, $S_i$ the set of strategies or choices available to that player and $S$ the set of all possible strategy profiles. In order to evaluate the profit or satisfaction of player $i$ with regards to a strategy profile we need to define for each player a payoff function $\$_i$ that takes a strategy profile $\sigma$ as input and outputs a real numerical value as a measure of desirability. We have $\$:S\rightarrow \mathbb{R}$ and $\$_i(\sigma_k)\geq\$_i(\sigma_l) \Leftrightarrow \sigma_k \succeq_i \sigma_l$. The question to answer is, what should rational players choose to do given that they have partial or complete information of the content of $S$ and the payoff functions $\$_i$? The main approach is to find a \emph{solution concept}, with the most famous one being the \emph{Nash Equilibrium}, where all players simply make the choice $s^i_j$ that is the best possible response to any configuration in $S/S_i$, i.e to any combination of strategies by their counter-parties. In situations where such equilibrium does not exist one needs to extend the game to allow for mixed (probabilistic) strategies where the players extend the sets $S_i$ to $\Delta(S_i)$, i.e the set of convex combinations of the $s^i_j$'s to acquire one.
Just as classical probability distributions extend  pure strategy games to mixed ones, quantum probabilities, operations and entanglement can extend the framework to outperform any classical setup.

\subsection{Quantum games}

A quantum game is defined by a set $\Gamma$ of objects and the relationships between them:
\begin{equation}\label{qg}
  \Gamma=\{\rho_{\mathcal{Q}},\mathcal{H}_{\mathcal{Q}}, n, S_{i}, \$_{i}\} \;\; \textrm{for}\;\; i=1, \cdots ,n
\end{equation}
where
$\mathcal{H}_{\mathcal{Q}}$ is the Hilbert
space of the composite quantum system, $\rho_{\mathcal{Q}}$ is the initial state of the game defined on $\mathcal{H}_{\mathcal{Q}}$, $n$ is the number of players, $S_{i}$ the set of available strategies of player $i$ and $\$_{i}$  the payoffs available to player $i$ for each game outcome. In our quantum game protocol the $m_i$ different pure strategies available to a player $i$ will be encoded in the basis states of an $m_i$-level quantum system $\rho_{\mathcal{Q}_{i}} \in \mathcal{H}_{{\mathcal{Q}_{i}}}$. With $n$ players we'll end up needing a initial quantum state $\rho_{\mathcal{Q}} \in \mathcal{H}_{\mathcal{Q}}=\mathcal{H}_{\mathcal{Q}_n}\otimes \mathcal{H}_{\mathcal{Q}_{n-1}}\cdots \otimes \mathcal{H}_{\mathcal{Q}_1}$ with $\textrm{dim}(\mathcal{H}_{\mathcal{Q}})= \prod_{i=1}^n  \textrm{dim}(\mathcal{H}_{\mathcal{Q}_i})$ to accommodate for all possible game outcomes \cite{review}.

The strategies are chosen and played by each player trough the application of a unitary operator $U_i \in S_i = S(m_i)$ on their own sub-systems, where the set of allowed quantum operations $S(m_i)$ is some subset of the special unitary group $\textrm{SU}(m_i)$. The general procedure of a quantum game consists of a transformation of the composite initial state trough local unitary operations by the players: $U_n\otimes U_{n-1} \otimes \cdots \otimes U_1:\mathcal{H}_{\mathcal{Q}} \rightarrow \mathcal{H}_{\mathcal{Q}}$. Followed by a measurement outcome, or in terms of pre-measurement reasoning, an expectation value: $\$: \mathcal{H}_{\mathcal{Q}} \rightarrow \mathbb{R}$.

\section{Quantum Kolkata restaurant problem}

This is a general form of a minority game \cite{kolkata1,Hayden,Chen}, where $n$ non-communicating agents (players), have to choose between  $m$ choices.
A payoff of $\$=1$ is payed out to the players that make \emph{unique} choices. Players making the same choice receive $\$=0$. The challenge is to come up with a strategy profile that maximizes the expected payoffs $E_i(\$)$ of all players $i$, and has the property of being a Nash equilibrium. In the absence of communication, in a classical framework, there is nothing else to do, but to randomize.

\subsection{Collective aim in the quantum case}
It has been shown for the case of three players and three choices in the quantum setting, starting with a GHZ-type state $|\psi_{in}\rangle=\frac{1}{\sqrt{3}}\left(|000\rangle+|111\rangle+|222\rangle\right) $, that shared entanglement and local SU(3) operations (by the players on their own subsystems) will lead to a expected payoff $E(\$)=\frac{2}{3}$. This is a 50\% increase compared to the classical payoff of $\frac{4}{9}$ reachable trough randomization. Although the details of the protocol can be found in \cite{puya}, it is instructive to jump back a couple steps.
Since we have three players with three allowed pure choices the Hilbert space we are dealing with is the space of three qutrit states; with a basis $B=\{|ijk\rangle \};\; i,j,k \in \{0,1,2\}$, each of which representing a post-measurement outcome of the game, where $i,j,k$ denotes the final choices of players $1,2,3$ respectively. We have $\textrm{span}(B) = \{\sum_{i,j,k =0}^2 a_{ijk}|ijk\rangle: i,j,k = 0,1,2 \; \textrm{and} \; a_{ijk} \in \mathbb{C}\}$ which with a normalization condition gives us the complete Hilbert space of the game. We can divide $B$ into subsets that are interesting from the point of view of the possible outcomes:

\begin{eqnarray}
  L\, &=& \{|000\rangle,|111\rangle,|222\rangle\}, \\
  G\, &=& \{|012\rangle,|120\rangle,|201\rangle,|021\rangle,|102\rangle,|210\rangle\},\\
  D_1 &=& \{|011\rangle,|022\rangle,|100\rangle,|122\rangle,|200\rangle,|211\rangle\},\\
  D_2 &=& \{|101\rangle,|202\rangle,|010\rangle,|212\rangle,|020\rangle,|121\rangle\},\\
  D_3 &=& \{|110\rangle,|220\rangle,|001\rangle,|221\rangle,|002\rangle,|112\rangle\},
\end{eqnarray}

where $L$ contains all states for which none the players 1,2,3 receive any payoff. It is thus a collective objective to avoid these states. $G$ contains all those states that returns a payoff $\$=1$ to the three of them and the sets $D_i$ contains the post-measurement states leads to a payoff $\$=1$ for player $i$ and $\$=0$ to players $\neq i$.
Thus the general goal of each player $i$ is to maximize the probability of the post-measurement outcome to be a state in $G_i=G \cup D_i$. Starting with an initial state $|\psi_{in}\rangle$, each player $i \in {1,2,3}$ applies an operator from its set of allowed strategies $S_i \subseteq \textrm{SU}(3)$, transforming it to its final state $|\psi_{fin}\rangle = U_1 \otimes U_2 \otimes U_3 |\psi_{in}\rangle$. The expected payoff $E(\$_i)$ of player $i$ is the probability  of the post-measurement outcome to be a state in $G_i$:

\begin{equation}\label{payoff}
    E(\$_i) = \sum_{|\xi\rangle \in G_i} \left|\langle \psi_{fin} |\xi\rangle \right|^2.
\end{equation}

Given that we have an initial state in $\textrm{span}(L)$ containing all states of the form $|\psi_{in}\rangle=\alpha|000\rangle+\beta|111\rangle+\gamma|222\rangle $ with $\alpha,\beta,\gamma \in \mathbb{C}$ (We will assume $0 \leq \alpha,\beta,\gamma \in \mathbb{R}$ later for simplicity), what is the rational aim of player $i$? First, note that all states in  $\textrm{span}(L)$ are unbiased with regards of change in player positions. We can assume that they don't even know which qutrit they control since that knowledge doesn't add any useful information for the choice of $U_i \in S_i$. Second, any choice of $U_i$ aimed at increasing the probability of post-measurement state to end up being in $D_i$ must due to the symmetry of the setup increase the probability of the state being in $D_{j\neq i \vee k} \cup D_{k\neq i \vee j}$ with 2:1. Third, an outcome in $G = G_1 \cap G_2 \cap G_3$ is as favorable as any outcome in $D_i$ for player $i$. It follows therefore that the players should aim for producing a state in $\textrm{span}(G)$ to the extent this is possible. Although it was shown in \cite{puya} that they fail to fully depart from $\textrm{span}(L)$, whereby they reach a maximum payoff of $E(\$)=\frac{2}{3}$ rater than $E(\$)=1$ for $\alpha = \beta = \gamma = \frac{1}{\sqrt{3}},$ with a final state of the following form:
\begin{multline}
\mid\psi_{fin}\rangle=\frac{1}{3}\left(|000\rangle+|012\rangle+|021\rangle+|102\rangle\right.+\\\left.|111\rangle+|120\rangle+|201\rangle+|210\rangle+|222\rangle\right).
\end{multline}

\subsection{Expected payoffs for initial states in span(\emph{L}) }
Due to the symmetries mentioned in the previous section the three players will have to chose a unitary operator $U$ that takes states from $\textrm{span}(L)$ to $\textrm{span}(L \cup G)$, without the possibility to favor any subset of $G$ (even if that possibility existed, that would only put a roof for the individual payoffs and any choice of subset other than the whole would decrease the expected payoff, due to the lack of coordination the choice of subset). This leads to the conclusion that there exists a $U$ (fixed up to a global phase) for any initial state $|\psi_{in}\rangle=\alpha|000\rangle+\beta|111\rangle+\gamma|222\rangle $, that maximizes the individual expected payoffs and is a Nash equilibrium solution, since any departure from this strategy will lead to a lower payoff. We have:

 \begin{equation}\label{payoff}
    E_{max}(\$) = \sum_{|\xi\rangle \in G} \left|\langle \psi_{fin}| U^{\dagger} \otimes U^{\dagger} \otimes U^{\dagger} |\xi\rangle \right|^2,
\end{equation}
simultaneously for all three of them. Figure 1 shows numerically calculated expected payoffs $E(\$)$ for $\alpha=\sin\vartheta\cos\varphi; \, \beta=\sin\vartheta\sin\varphi; \, \gamma=\cos\vartheta$ where $\varphi= \frac{\pi}{40}M,\vartheta= \frac{\pi}{40}N$ and $M,N = 1,2, \cdots, 20$. A total of 400 optimizations with equally many different  associated operators $U_{MN}$. We see that the expected payoff is maximized for $\varphi=\frac{\pi}{4}$  and $\vartheta=\cos^{-1}\frac{1}{\sqrt{3}}$, where the initial state is maximally entangled, and falls off towards the classical expected payoff as the entanglement decreases.

\begin{figure}
%\centering
  \includegraphics[scale=0.75]{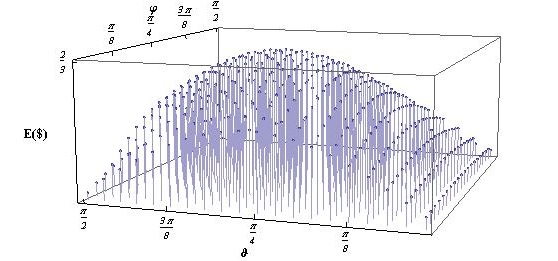}\\
  \caption{Payoffs associated with optimal strategies in a three player game with a variable initial state. The expected payoff $E(\$)$ decreases as the level of entanglement decreases.}\label{b}
\end{figure}

\section{Conclusions}

The ambition of self-maximization in the studied Kolkata restaurant problem leads individual players to act in such way that the collective good is maximized. The game is symmetric with regards to permutations of player positions which guides the participants to aim for a set of outcomes that favors them all. The expected payoff reachable trough local operations changes with the level of entanglement in the initial state.

\subsection*{Acknowledgements:}
The work was supported by the Swedish Research Council (VR).



\bibliographystyle{aipproc}   % if natbib is available

%\bibliography{sample}


%%%%%%%%%%%%%%%%%%%%%%%%%%%%%%%%%%%%%%%%%%%
\IfFileExists{\jobname.bbl}{}
 {\typeout{}
  \typeout{******************************************}
  \typeout{** Please run "bibtex \jobname" to optain}
  \typeout{** the bibliography and then re-run LaTeX}
  \typeout{** twice to fix the references!}
  \typeout{******************************************}
  \typeout{}
 }

%%%%%%%%%%%%%%%%%%%%%%%%%%%%%%%%%%%%%%%%%%%
%% The following lines show an example how to produce a bibliography
%% without the help of the BibTeX program. This could be used instead
%% of the above.
%%%%%%%%%%%%%%%%%%%%%%%%%%%%%%%%%%%%%%%%%%%

\end{document}